\documentclass[aps,prb,twocolumn,showpacs,amsmath,amssymb]{revtex4-1}
\usepackage{dcolumn}% Align table columns on decimal point
\usepackage{hyperref}
\usepackage{color}
\usepackage{amssymb}
\usepackage {graphicx}
\usepackage {sidecap}
\usepackage{sidecap}
\usepackage{graphicx}
\usepackage{epstopdf}
\usepackage[normalem]{ulem}
 \usepackage{gensymb}
\newcommand{\beq}{\begin{equation}}
\newcommand{\eeq}{\end{equation}}
\newcommand{\bea}{\begin{eqnarray}}
\newcommand{\eea}{\end{eqnarray}}
\newcommand{\be}{\begin{equation}}
\newcommand{\ee}{\end{equation}}

\renewcommand{\k}{\boldsymbol{k}}

\newcommand{\q}{\boldsymbol{q}}

\newcommand{\kq}{\boldsymbol{k}+\boldsymbol{q}}
\newcommand{\bl}{\begin{aligned}}
\newcommand{\el}{\end{aligned}}
\newcommand{\mi}{\rm i}

\newcommand{\ba}{\begin{align}}
\newcommand{\ea}{\end{align}}
\graphicspath{{Figs/}}
\begin{document}

\title{Spin-orbit coupling, minimal model and potential Cooper-pairing from repulsion in BiS$_2$-superconductors}
\author{Sergio Cobo-Lopez$^{1,2}$}
\email{Present address: Departament d’Enginyeria Química, Universitat Rovira i Virgili, 43007 Tarragona, Catalonia, Spain}
\author{M. S. Bahramy$^{3,4}$}
\author{Ryotaro  Arita$^{3,4}$}
\author{Alireza Akbari$^{1,5,6}$}
\email{alireza@apctp.org}

\author{Ilya Eremin$^{2,7}$}
\email{Ilya.Eremin@rub.de}

\affiliation{
$^{1}$Asia Pacific Center for Theoretical Physics (APCTP), Pohang, Gyeongbuk, 790-784, Korea\\
$^2$Institut f\"ur Theoretische Physik III, Ruhr-Universit\"at Bochum, D-44801 Bochum, Germany\\
$^3$Department of Applied Physics, The University of Tokyo, Tokyo 113-8656, Japan \\
$^4$RIKEN center for Emergent Matter Science (CEMS), Wako 351-0198, Japan\\
$^{5}$Department of Physics, POSTECH, Pohang, Gyeongbuk 790-784, Korea\\
$^{6}$Max Planck POSTECH Center for Complex Phase Materials, POSTECH, Pohang 790-784, Korea \\
$^7$ National University of Science and Technology “MISiS”, 119049 Moscow, Russian Federation
}

\date{\today}

\begin{abstract}
We develop the realistic minimal electronic model for recently discovered BiS$_2$ superconductors including the spin-orbit coupling based on a first-principles band structure calculations. Due to strong spin-orbit coupling, characteristic for the Bi-based systems, the tight-binding low-energy model necessarily includes $p_x$, $p_y$, and $p_z$ orbitals. We analyze a potential Cooper-pairing instability from purely repulsive interaction for the moderate electronic correlations using the so-called leading angular harmonics approximation (LAHA). For small and intermediate doping concentrations we find the dominant instabilities to be $d_{x^2-y^2}$-wave, and $s_{\pm}$-wave symmetries, respectively. At the same time, in the absence of the sizable spin fluctuations the intra and interband Coulomb repulsion are of the same strength, which yields the strongly anisotropic behaviour of the superconducting gaps on the Fermi surface in agreement with recent ARPES findings. In addition, we find that the Fermi surface topology for BiS$_2$ layered systems at large electron doping can resembles the doped iron-based pnictide superconductors with electron and hole Fermi surfaces with sufficient nesting between them. This could provide further boost to increase $T_c$ in these systems.
\end{abstract}


\maketitle

%%%%%%%%%%%%%%%%%%%%%%%%%%%%%%%%%%%%%%%%%%%%
\section{Introduction}
Recent discovery of BiS$_2$-layered superconductors with T$_c$ up to
10K has attracted significant attention due to striking similarities of their crystal structure with cuprate and iron-based high-T$_c$ superconductors~\cite{Mizuguchi2012,Singh2012,Mizuguchi2,Demura,Xing,Jha,Lin2013,Yazici2013,Deguchi2013,Biswas2013,YKLi2014}.
According to the density functional theory (DFT)  calculations, the parent compounds of the BiS$_2$-based superconductors are semiconductors and the metallic behavior is achieved by electron doping of the conduction
band, which mainly consists of Bi 6$p$ orbitals~\cite{Mizuguchi2012,Usui2012}.

Although the origin of superconductivity is not yet clear in these compounds, it was attributed to the electron-phonon interaction~\cite{Wan2013,Li2013,Yildirim2013}. This seems reasonable in view of the low superconducting transition temperature and weaker electronic correlations of $p$-orbitals than in their 3$d$-counterparts. However, in the recent neutron scattering experiment, the observed almost unchanged low-energy modes indicated that the electron phonon coupling could be weaker than expected~\cite{Lee2013}. In addition, the large ratio $2\Delta/T_c$ may suggest that the pairing mechanism is unconventional~\cite{SLi2013,Liu2014}. In addition, no isotope effect on superconducting transition temperature was found\cite{hoshi2017}.

Very recently, using `ab-initio' approach the strength of the electron-phonon coupling in LaO$_{0.5}$F$_{0.5}$BiS$_2$ was estimated to be $\lambda <  0.5$~\cite{Morice2017}, which is too low to explain the superconducting transition temperature in this system. 
Electron-electron correlations can, in principle, be also responsible for the Copper pairing and there were several theoretical studies about possible
pairing symmetries in these new superconductors arising due to repulsive interactions~\cite{Zhou2013,Liang2014,Matrins2013,Yang2013}.  However, as the nominal compositions of the superconducting materials referred to the significant electron doping~\cite{Mizuguchi2,YKLi2014}, all these previous
studies concentrated on the high electron doping region where
the electronic structure was featured with large Fermi surfaces in close
vicinity to Van Hove singularity. However, very recently, it was reported that the actual electronic filling in these systems is much lower than the nominal one and  there are only two small
electron pockets around the $X$ points of the Brillouin Zone~\cite{Zeng2014,Ye2014}  corresponding to $x=0.22$~\cite{Sugimoto2015}. 
Most importantly, the recent observation of the strongly anisotropic superconducting gaps on the electron pockets in  NdO$_{0.71}$F$_{0.29}$BiS$_2$ using laser angle resolved photoemission spectroscopy (laser ARPES) suggests that the pairing mechanism for this system is likely an unconventional one and was attributed to be a result of the multiple paring interactions\cite{Ota2017}.

Here, based on the fully relativistic ab-initio band structure calculations, we develop the realistic minimal electronic model for recently discovered BiS$_2$ superconductors including the spin-orbit coupling, relevant for Bi-based systems. We show that the tight-binding low-energy model necessarily includes all three $p$-orbitals, $p_x$, $p_y$, and $p_z$. Using angular harmonics approximation (LAHA)\cite{ahn2014} we analyze a potential Cooper-pairing instability from purely repulsive interaction concentrating on the case of weak to intermediate electronic correlations strength for small and intermediate doping levels. Similar to the previous spin fluctuations based studies\cite{Usui2012,Dai2015} we find global $d_{x^2-y^2}$ (B$_{1g}$)-wave, and anisotropic $s$-wave (A$_{1g}$) symmetries, respectively, to be the dominant symmetries. However, our results show that for weak correlations, i.e. when the intraband and the interband repulsion are of similar strengths, the gap cannot be described by the single harmonics of the corresponding functions of the $A_{1g}$ or $B_{1g}$ irreducible representations typical for the case of the strong spin fluctuations. Instead, each gap acquires a strong angular dependence in the form of the multiple harmonics on each of the Fermi surface pockets allowed by the global symmetry. This is a consequence of the fact that the electronic system reduces the net repulsion. We anaylze the form of these harmonics and show that they are in agreement with recent experimental ARPES observation\cite{Ota2017}. We further investigate the Fermi surface topology for BiS$_2$ layered systems at large electron doping and show its similarity with the doped iron-based pnictide superconductors with electron and hole Fermi surfaces with sufficient nesting between them. This could provide further boost to increase $T_c$ in these systems.

%

%%%%%%%%%%%%%%%%%%%%%%%%%%%%%%%%%%
\section{Tight-binding model}
\begin{figure}[!t]
\label{fig1}
\includegraphics[width=1.01\columnwidth]{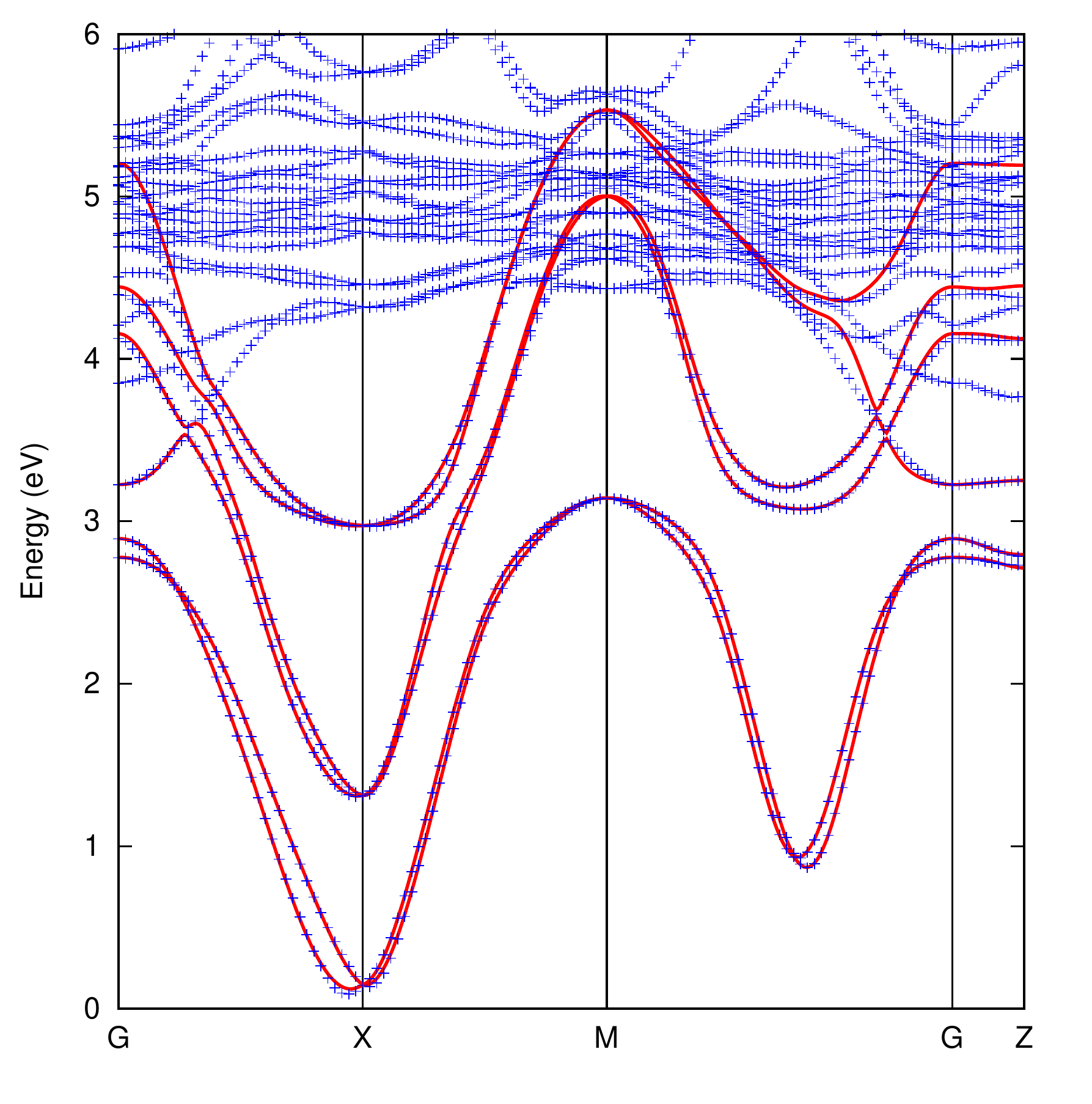}
\caption{
(Color online) Band structure of LaOBiS$_2$ as obtained directly form a DFT calculation (indicated by $+$ symbols) and as interpolated using atomic-like Wannier functions centered at Bi sites. 
For the latter, $p_{x,y,z}$ orbitals of Bi are used as the projection centers. Note that our basis is rotated by 45$^{\degree}$ with respect to the one, employed in Ref.~\cite{Usui2012}.
} 
\label{fig1_dft}
\end{figure}

The fully relativistic band structure calculation of the LaOBiS$_2$ system was
performed  within DFT using the Perdew-Burke-Ernzerhof exchange-correlation functional as implemented in WIEN2K program~\cite{wien2k}.  The muffin-tin radius of each atom $R_{\text MT}$ was chosen such that its product with the maximum modulus of reciprocal vectors $K_{\text max}$ become $R_{\text MT}K_{\text max}=7.0$. The lattice parameters were taken from Ref.~\cite{Tanryverdiev1995} and the corresponding Brillouin zone was sampled using a $12\time12\times4$~$k-$mesh. 
From this band calculation, as shown in Fig.\ref{fig1_dft}, we constructed a 12-band tight binding (TB) model using the maximally localized
Wannier functions~\cite{souza2001,mostofi2008,kunes2010} with Bi 6$p$  ($p_x$, $p_y$, and $p_z$) orbitals  as projection centers.  
The resulting  TB model was further reduced to a 3-band model 
%
%-------------------------------------------------------------------------
\be
\hat{\mathcal{H}}_0=
\sum_{ij}\sum_{\nu\nu'\sigma}
t_{ij;\nu \nu'}
c_{i\nu\sigma}^\dagger c_{j\nu'\sigma}^{}
+\sum_{i\nu\sigma}\varepsilon_\nu c_{i\nu\sigma}^\dagger c_{i\nu\sigma}^{}, % n_{i\mu\sigma},
\ee
%-------------------------------------------------------------------------
%
%
plus spin-orbit coupling by ignoring the inter-layer hopping between the BiS$_2$ layers. 
Here, $t_{ij;\nu \nu'}=t(x_i-x_j, y_i-y_j;\nu \nu')$ are the nearest and next-nearest neighbor hopping parameters obtained from projecting the results of ab initio calculations. 
Their  values are shown in Table I where $\nu$, and $\nu'$ refer to  the orbital indices. 
Thus, in the  %Fourier transform 
Momentum space
 the Hamiltonian can be written as
%
%
%-------------------------------------------------------------------------
\begin{equation}
\hat{\mathcal{H}}_0 
= \sum_{\mathbf{ k}\nu \nu' \sigma} T_{\nu\nu'}
(\mathbf{ k})
c^\dagger_{\mathbf{k} \nu \sigma} c^{\phantom{\dagger}}_{\mathbf{ k}\nu' \sigma},
\end{equation}
%-------------------------------------------------------------------------
%
with the hopping integrals
%-------------------------------------------------------------------------
\bea
\bl\nonumber
T_{\nu \nu} 
=& t_{0,0}^{\nu} + 2t_{1,0}^{\nu} \cos k_x + 2t_{0,1}^{\nu} \cos k_y 
\\
& + 2t_{1,1}^{\nu} \cos (k_{x} + k_{y}) + 2t_{1,-1}^{\nu} \cos (k_x - k_y), 
\\
T_{\nu \nu'}
=&
T_{\nu' \nu}
=
t_{0,0}^{\nu \nu'} +2t_{1,0}^{\nu \nu'} \cos k_x + 2t_{0,1}^{\nu \nu'} \cos k_y  
\\
&\hspace{0.8cm}
+ 2t_{1,1}^{\nu \nu'} \cos (k_{x} + k_{y}) + 2t_{1,-1}^{\nu \nu'} \cos (k_x - k_y).
\el
\eea
%-------------------------------------------------------------------------
Here $t_{x,y}^{\nu}$ and $t_{x,y}^{\nu \nu'}$ denote intra- and interorbital electron hopping,
 as determined in Table I.
 \\
 
 On top of that, the on-site spin-orbit (SO) coupling Hamiltonian, $H_{SO} = \lambda ~  {\bf L} \cdot {\bf S}$, written for the bismuth $p$-orbitals %($p_x$, $p_y$, $p_z$) 
 have the following form:
%--------------------------------------------------------------------
\begin{equation}
 \mathcal{H}_{SO} = \frac{\lambda}{2}
 \left( \begin{array}{ccc@|ccc}
0 & - \mi & 0 & 0 & 0 & 1 \\
\mi & 0 & 0 & 0 & 0 & -\mi \\
0 & 0 & 0 & -1 & \mi & 0 \\
$---$ & $---$ &$---$ & $---$ & $---$& $---$ \\
 0 & 0 & -1 & 0 & \mi & 0 \\
0 & 0 & -\mi & -\mi & 0 & 0 \\
1 & \mi & 0 & 0 & 0 & 0
\end{array} \right),
\end{equation}
 %--------------------------------------------------------------------------
 %
%in the rest of the paper,
hereafter,  we set $\lambda=0.874$eV as a fitting parameter to match the results of the effective model to the DFT band structure. %   (see Fig.\ref{fig1_dft} \& Fig.\ref{fig2}(a)).
The full electronic structure of the resulting tight-binding Hamiltonian is shown in Fig.\ref{fig2}(a). For the low filling the band structure can be indeed reproduced roughly by taking account only the $p_x$ and $p_y$ orbitals and the parameters we find are similar to the ones found previously~\cite{Usui2012}.
%At the same
We should note that the proper inclusion of the spin-orbit coupling does require to involve  the $p_z$-orbital  into consideration. This is valid  even for the small fillings that always  there is  an admixture of the $p_z$ orbitals to the lowest band 
due to spin-orbit coupling. %, which would otherwise be purely $p_x/p_y$ origin. 

%
%+++++++++++++++++++++++++++++++   figure   ++++++++++++++++++++++++++++++++++++++++++++++++++
%++++++++++++++++++++++++++++++++++++++++++++++++++++++++++++++++++++++++++++++++++++++++++
%

For the small fillings the Fermi surface of the BiS$_2$-layer consists of the two electron pockets centered near $(\pm \pi,0)$ and $(0, \pm \pi)$ points of the BZ, which resembles some of the iron-based electron-doped chalcogenide superconductors. Upon doping the electron pockets become bigger and at $x=0.5$ the system undergoes the transition from the small electronic Fermi surface pockets to the large Fermi surface, similar to Ref.~\cite{Usui2012}. 
Due to spin-orbit coupling the contribution of $p_z$-orbital to the conducting band increases and becomes quite significant for $x>0.5$. Studying the evolution of the electron structure for larger doping we observe that for the doping range $n>1.5$,
 the Fermi surface topology resembles strongly the one found in the iron pnictide superconductors (see Fig.\ref{fig2}(c)).
 That corresponds to the weakly nested electron and hole Fermi surfaces centered near the corresponding points of the BZ. 
 Such a nesting tends to boost the inter-band electron-electron scattering and favor non-phononic mechanism of superconductivity.

%
%+++++++++++++++++++++++++++++++   figure   ++++++++++++++++++++++++++++++++++++++++++++++++++
\begin{figure}[b]
\label{fig2}
\includegraphics[width=1.04\columnwidth]{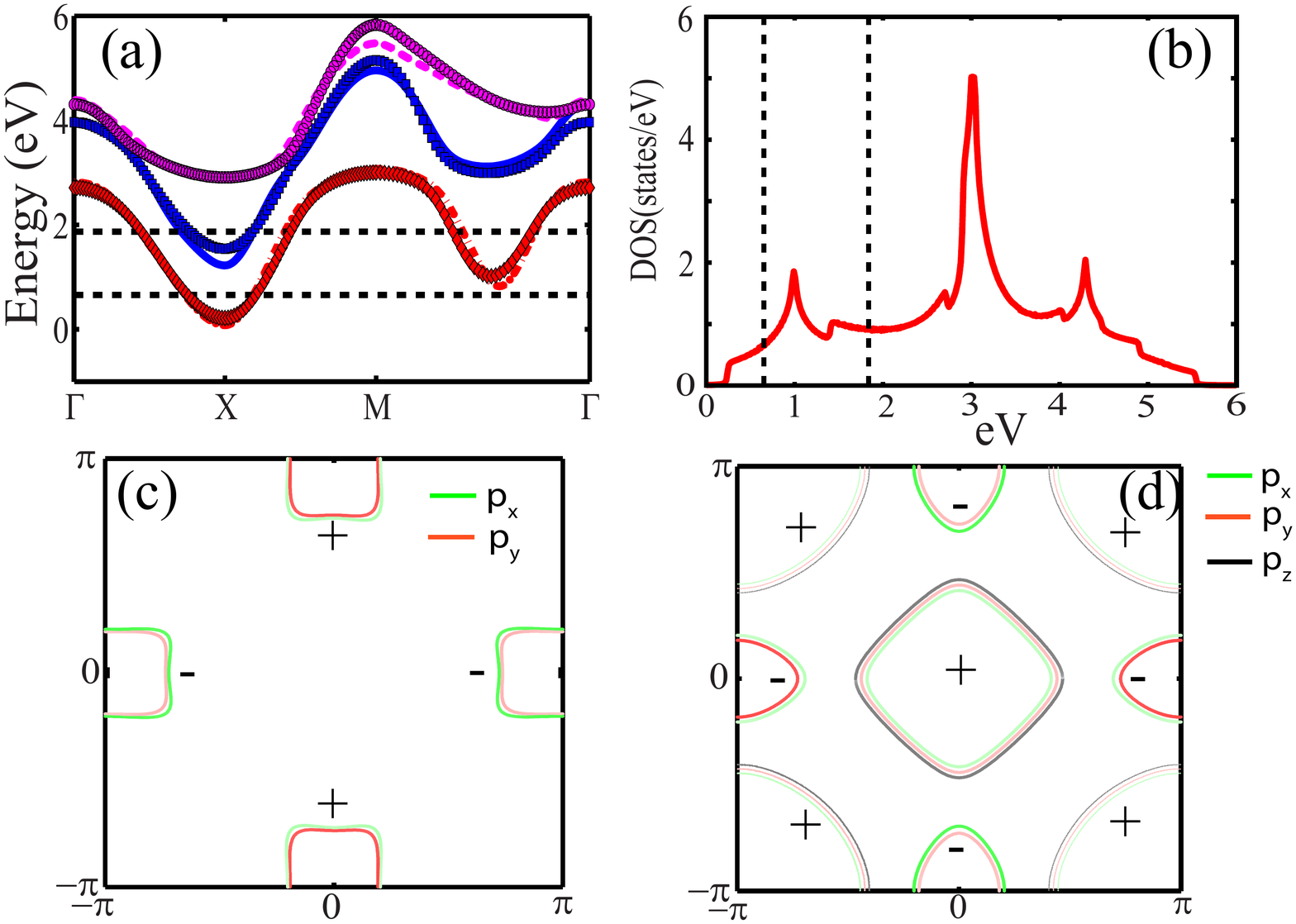}
\caption{(Color online) Electronic structure of the BiS$_2$-layered superconductors. (a) shows the result of tight-binding parametrization of DFT bands (dashed curves) using the three orbitals $p_x$, $p_y$, and $p_z$ (solid curves) fit, including spin-orbit coupling. The corresponding density of states (b) with two representative chemical potential positions, and the resulting Fermi surfaces (c-d)  are also shown. The $\pm$ signs in (c) and (d) refer to the phase structures of the nodeless $d_{x^2-y^2}$-wave and $s^\pm$- wave gaps. The orbital contribution for each Fermi surface is presented with slight offset for clarity.} \label{fig2}
\end{figure}
%++++++++++++++++++++++++++++++++++++++++++++++++++++++++++++++++++++++++++++++++++++++++++
%

%
%++++++++++++++++++++++++++++++++++++   Table   +++++++++++++++++++++++++++++++++++++++++++++
\begin{center}
%\begin{SCtable*}
\begin{table*}
\label{table 1}
\begin{tabular}{@{}|c|ccccccccc|@{}} \hline
\multicolumn{1}{|c|}{($\nu,\nu'$)} & \multicolumn{9}{c|}{[$\Delta x, \Delta y$]}  \\\cline{2-10} 
\multicolumn{1}{|c|}{} & [0,0] & [1,0] & [0,1] & [-1,0] & [0,-1] & [1,-1] & [-1,1] & [1,1] & [-1,-1] \\\hline
$(1,1)$ & 0 & -82.5 & -298 & -82.5 & -298 & 457 & 457 & 459 & 459   \\
$(2,2)$ & 0 & -298 & -82.5 & -298 & -82.5 & 457 & 457 & 459 & 459  \\
$(3,3)$ & 182 & $5.78$ & $5.78$ & $5.78$ & $5.78$ & 55 & 55 & 65 & 65.5 \\
$(1,2)$ & $-3.01$ & $0.513$ & $0.562$ & $0.711$ & $0.662$ & -385 & -385 & 386 & 385  \\
$(1,3)$ & $7.78$ & $-0.337$ & $7.19$ & $-0.393$ & $7.20$ & $-5.27$ & $-5.38$ & $0.269$ & $0.175$  \\
$(2,3)$ & $-7.78$ & $-7.72$ &  $0.389$ & $-7.19$ & $0.334$ & $5.27$ & $5.38$ & $-0.177$ & $-0.272$ \\\hline
\end{tabular}
\caption{Hopping parameters $t(x_i-x_j, y_i-y_j; \nu \nu')=t(\Delta x, \Delta y; \nu \nu')$ (in meV) for the simplified two-dimensional model. Note that t$(\Delta x, \Delta y;\nu \nu')$ =t(-$\Delta x, -\Delta y; \nu \nu'$) and t$(\Delta x, \Delta y; \nu \nu')$ =t($\Delta x, \Delta y; \nu' \nu$).
}
%\end{SCtable*}
\end{table*}
\end{center}
%+++++++++++++++++++++++++++++++++++++++++++++++++++++++++++++++++++++++++++++++++++++++++
%+++++++++++++++++++++++++++++++++++++++++++++++++++++++++++++++++++++++++++++++++++++++++

%%%%%% Section %%%%%%%%%%%%%%%%%%%%%%%
\section{Lindhard response function}
As mentioned in the introduction, the origin of superconductivity in BiS$_2$ layers is still debated. Nevertheless, given the similarity of the BiS$_2$-systems and the iron-based superconductors regarding crystal structure and Fermi surface topology, it is tempting to apply the concept of Cooper-pairing from repulsion to these systems as well, given much weaker correlations strength in Bi-based materials. Previously, assuming spin fluctuations scenario and the presence of the disconnected Fermi surfaces for the small electron doping the $d_{x^2-y^2}$-wave symmetry for the BiS$_2$ systems was obtained~\cite{Usui2012,Dai2015}. At the same time, recent ARPES experiments point towards strongly anisotropic superconducting gap\cite{Ota2017}, which would not agree with the simple spin-fluctuation mediated scenario with strong interband repulsion. Here, we concentrate on the more realistic situation of the weak spin fluctuations and analyze the potential Cooper-pairing instabilities for the case of disconnected Fermi surface pockets using the leading angular harmonics approximation (LAHA), developed previously and specially designed for the systems when the spin fluctuations are either intermediate or weak.  \cite{Maiti2011a,Maiti2011b,ahn2014}.  
To search for the potential instabilities of the electron system, we perform the calculations of the Lindhard response function, which gives information on the potential instabilities of electronic system. For the multiorbital systems  
Based on the Lindhard spin response function theory, the bare   spin susceptibility is defined by
 %--------------------------------------------------------------------------
\begin{equation} \label{suscep}
(\chi^{s s'})^{\nu_3\nu_4}_{\nu_1\nu_2}(\boldsymbol{q})=
\Big\langle 
T \hat{S}^{s}_{\nu_1\nu_2}(\boldsymbol{q},\tau) \hat{S}^{s'}_{\nu_3\nu_4}(-\boldsymbol{q},0) 
\Big\rangle,
\end{equation}
 %--------------------------------------------------------------------------
where  $\nu_i$ runs over orbitals, and $\hat{S}^{s}$ %and $\hat{S}^{s'}$ 
are the spin operators  given by
 %--------------------------------------------------------------------------
\begin{equation} \label{spin}
\hat{S}^{s}_{\mu\nu}(\boldsymbol{q},\tau)=
\frac{1}{2}
\sum_{\substack{\k \sigma \sigma'}}
c^{\dagger}_{\k\mu\sigma}(\tau)
\hat{\sigma}^{s}_{\sigma \sigma'} 
c^{}_{\kq\nu\sigma'}(\tau),
\end{equation}
 %--------------------------------------------------------------------------
%where $a$ and $b$ represent the  orbitals and $\gamma$ and $\delta$ are  labeled the spin indices of the Pauli matrix $\hat{\sigma}^{s}$. 
where
$\hat{\sigma}^s$ are Pauli matrices.
Therefore, by applying Wick's theorem and neglecting contractions leading to $\q=\boldsymbol{0}$,   
one can find that the  different components of the bare spin susceptibility, $\chi^{ss'}_0$,  in the frequency domain is %given by
%------------------------------------------------------------------------------------------------------------------------------------------------------------------------------------------------------------------------------------------------
\be
\bl %\nonumber
&
(\chi^{s s'}_0)^{\nu_2 \nu_1}_{\nu_3 \nu_4}(\boldsymbol{q},\mi \Omega)
=
\frac{-T}{4 N}
\sum_{\substack{\k, \mi \omega_n \\ \gamma \delta \gamma' \delta'}} 
\!\!
\hat{\sigma}^{s}_{\gamma \delta} \hat{\sigma}^{s'}_{\gamma' \delta'} 
%\\
%&\hspace{1cm}
%
G_{\nu_4\nu_1,\delta' \gamma}^{\k,\mi \omega_n} 
G_{\nu_2\nu_3,\delta \gamma'} 
^{\kq,\mi \omega_n^{\prime}},
%^{(\kq,\mi \omega_n + \mi \Omega)},
\el
\ee
%-------------------------------------------------------------------------
%
with the Green's functions  Introduced by
%
%------------------------------------------------------------------------------------------------------------------------------------------------------------------------------------------------------------------------------------------------
\be
G_{\nu\nu',\sigma\sigma' }^{\k,\mi \omega_n}=
-\int^{\beta}_{0} d \tau e^{\mi \omega_n \tau} 
\Big\langle 
T_{\tau}~
c_{\k \nu \sigma} (\tau) 
c^{\dagger}_{\k \nu' \sigma'} (0) 
\Big\rangle,
\ee
%------------------------------------------------------------------------------------------------------------------------------------------------------------------------------------------------------------------------------------------------
%
here we set $\mi \omega'= \mi\omega_n + \mi \Omega $. 
Since the Green's functions are not diagonal in the orbital basis it is convenient to move to the band representation, and find  %, i.e.
% 
%------------------------------------------------------------------------------------------------------------------------------------------------------------------------------------------------------------------------------------------------
\begin{equation}
G_{\nu\nu',\sigma \sigma'}^{\k,\tau}
=\sum_{\mu} 
\zeta^{\mu}_{\k \nu \sigma}
\zeta^{\mu *}_{\k \nu' \sigma'} 
G_{\mu} ^{\k, \tau},
\end{equation}
 %------------------------------------------------------------------------------------------------------------------------------------------------------------------------------------------------------------------------------------------------
%
where $\mu$ being the band index, and  the matrix elements $\zeta^{\mu}_{\k \nu \sigma}$, are connecting the band to the orbital basis:
 %--------------------------------------------------------------------------
%\begin{equation}
$
c_{\k \nu \sigma}(\tau)=\sum_{\mu} \zeta^{\mu}_{\k \nu \sigma} b_{k \mu} (\tau)
$.
%\end{equation}
%--------------------------------------------------------------------------
%  Introducing these Green's functions and 
By performing the Matsubara frequency sum over  $\omega_n$,
%$\mi \omega_n=\omega+\mi \eta$  
yields
%
%------------------------------------------------------------------------------------------------------------------------------------------------------------------------------------------------------------------------------------------------
\begin{equation}
\bl
&
(\chi^{ss'}_0)^{\nu_2 \nu_1}_{\nu_3 \nu_4}(\boldsymbol{q},\omega)=-\frac{T }{4N}\sum_{
%}\sum_{
\substack{\k, \mu \mu' \\ \gamma,\delta,\gamma',\delta'}} 
\hat{\sigma}^{s}_{\gamma \delta} \hat{\sigma}^{s'}_{\gamma' \delta'} 
\times
\\  %\nonumber
& 
\Big( 
\zeta^{\mu}_{\k \nu_4 \delta'}
\zeta^{\mu *}_{\k \nu_1 \gamma} 
\zeta^{\mu'}_{\kq \nu_2 \delta}
\zeta^{\mu' *}_{\kq \nu_3 \gamma'} 
\Big)
\frac{f(E_{\kq,\mu'})-f(E_{\k,\mu})}{ \omega+\mi 0^{+} + E_{\kq,\mu'}-E_{\k,\mu}},
\el
\end{equation}
 %------------------------------------------------------------------------------------------------------------------------------------------------------------------------------------------------------------------------------------------------
%
where $f(\ldots)$ are Fermi weight functions. 
The straightforward  calculations based on  the Pauli matrices relation: 
\[\hat{\sigma}^{+}_{\gamma \delta}\hat{\sigma}^{-}_{\gamma' \delta'}=\left\{ \begin{array}{lcc}
4 &   $if$  & \gamma=\delta' %=\uparrow
\neq 
\gamma'=\delta %=\downarrow 
\\
\\ 0 &   & $otherwise$ \\
\end{array}
\right.\]
and
\[\hat{\sigma}^{z}_{\gamma \delta}\hat{\sigma}^{z}_{\gamma' \delta'}=\left\{ \begin{array}{lcc}
1 &   $if$  & \gamma=\delta=\gamma'=\delta' \\
-1 &   $if$  & \gamma=\delta=-\gamma'=-\delta' \\
0 &   & $otherwise$ \\
\end{array}
\right.\]
defined the transverse susceptibility as follows 
%------------------------------------------------------------------------------------------------------------------------------------------------------------------------------------------------------------------------------------------------
\begin{equation}
\bl
&
(\chi^{\pm}_0)^{\nu_2 \nu_1}_{\nu_3 \nu_4}(\boldsymbol{q},\omega)=\frac{-T }{4N}
 \sum_{\k \mu \mu'} 
 \\&
\Big( 
  \zeta^{\mu }_{\k \nu_4 \uparrow}
  \zeta^{\mu  *}_{\k \nu_1 \uparrow} 
  \zeta^{\mu'}_{\kq \nu_2 \downarrow}
  \zeta^{\mu' *}_{\kq \nu_3 \downarrow}
  \Big)
 \frac{f(E_{\kq,\mu'})-f(E_{\k,\mu})}{ \omega+\mi 0^{+} + E_{\kq,\mu'}-E_{\k,\mu}} ,
\el
\end{equation}
 %------------------------------------------------------------------------------------------------------------------------------------------------------------------------------------------------------------------------------------------------
and by considering $ \tilde{\sigma}=-\sigma$, we find 
%------------------------------------------------------------------------------------------------------------------------------------------------------------------------------------------------------------------------------------------------
\begin{equation}
\bl
&(\chi^{zz})^{\nu_2 \nu_1}_{\nu_3 \nu_4}
(\boldsymbol{q},\omega)=
\\
%
%&
&\hspace{1cm} 
\frac{-T }{4N}
 \sum_{\k \mu\mu'  \sigma }
 \Big(
 \zeta^{\mu  }_{\k \nu_4 \sigma}
 \zeta^{\mu   *}_{\k \nu_1 \sigma} 
 \zeta^{\mu  '}_{\kq \nu_2 \sigma}
 \zeta^{\mu ' *}_{\kq \nu_3 \sigma}
\\
&\hspace{1cm} 
-
\zeta^{\mu }_{\k \nu_4 \sigma}
\zeta^{\mu  *}_{\k \nu_1 \tilde{\sigma}} 
\zeta^{\mu '}_{\kq \nu_2 \tilde{\sigma}}
\zeta^{\mu ' *}_{\kq \nu_3 \sigma}
\Big)
\times
\\
&\hspace{3cm} 
\frac{f(E_{\kq,\mu'})-f(E_{\k,\mu})}{ \omega+\mi 0^{+} + E_{\kq,\mu'}-E_{\k,\mu}} ,
\el
\end{equation}
 %------------------------------------------------------------------------------------------------------------------------------------------------------------------------------------------------------------------------------------------------
for the longitudinal susceptibility. 

\begin{figure}[t]
\includegraphics[width=1.0\linewidth]{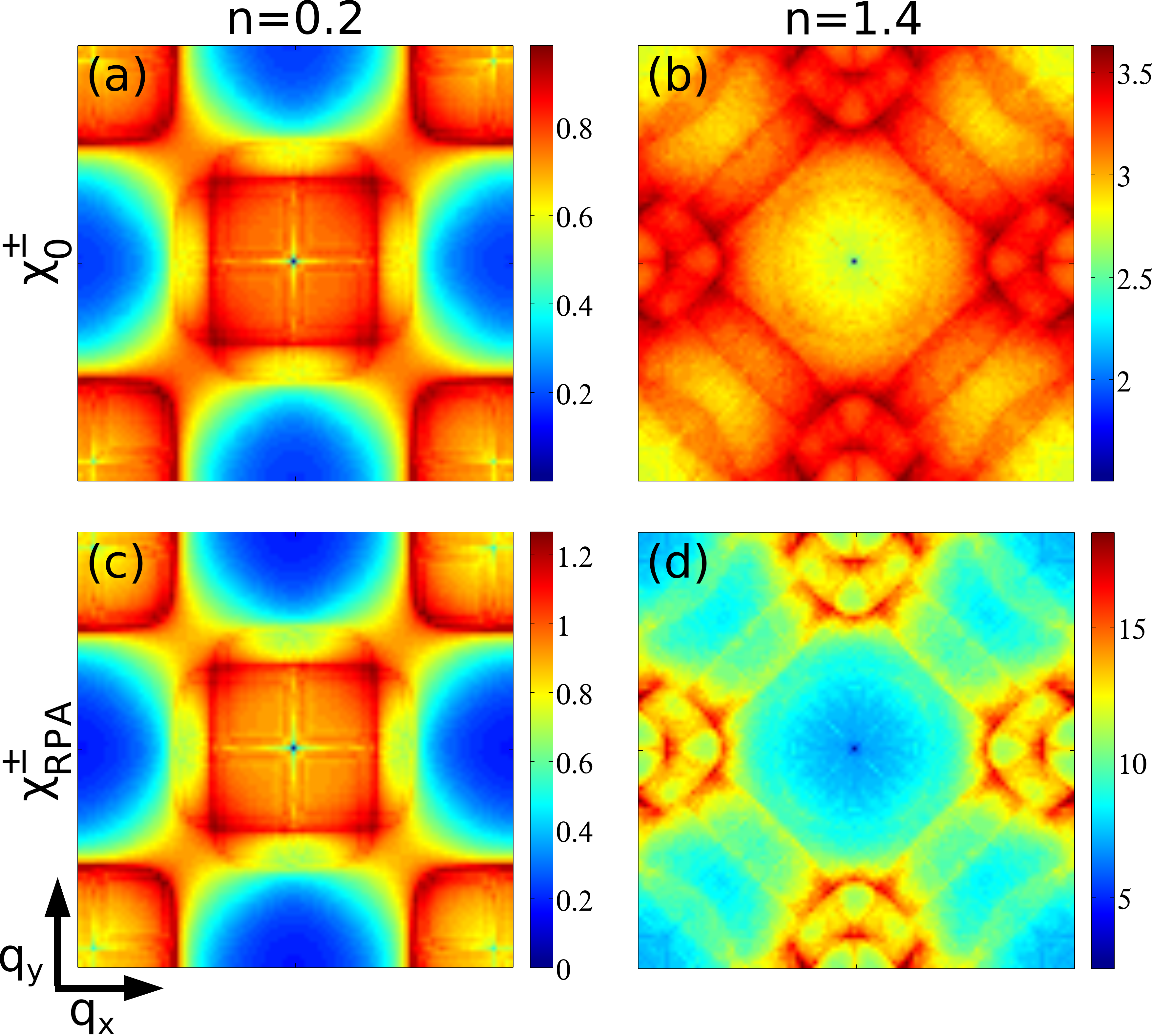}
\caption{(Color online)
(a) and (b) Bare  transversal susceptibilities in ${\rm states}/eV$ for doping levels $n=0.2$ and $n=1.4$, respectively. Susceptibility values are overall higher for $n=1.4$ due to a larger scattering rate consistent with the FS topology.  (c, d) Corresponding RPA transversal susceptibilities for the same doping levels. The values of the Hubbard Hamiltonian parameters are given by:  $U=0.6$, $V=0.42$, $J=J'=0.14$, which corresponds to the moderate level of the correlations. 
The peaks in the bare susceptibility are enhanced within the RPA only for the larger value of $n$. If the incommensurate peak at $(\pi/3,\pi/3)$ and the one close to $(\pi,\pi)$ for $n=0.2$ are only slightly enhanced,  the peaks close to the antiferromagnetic wave-vector $Q=(\pi,0)$ for $n=1.4$ are enhanced significantly due to the fact the the pockets have a different character. While the pockets near the $\Gamma$ and M pockets are hole like, the pockets near $X$ and $Y$-pints are electron like. Momentum ranges are given by $(-\pi\le q_{x/y}\le \pi)$.} 
\label{fig3}
\end{figure}

The results of the calculations for the two representative concentrations of the model are present in Fig\ref{fig3} (a),(b). For $n=0.2$ and $n=1.4$ the effect of the spin orbit-coupling is an introduction of the weak easy-plane anisotropy, i.e. $\chi^{+-}>\chi^{zz}$, which appears to be an overall prefactor. Therefore we discuss only the behaviour of $\chi^{+-}$ and the situation for $\chi^{zz}$ is similar. For $n=0.2$ we clearly observe the intraband (small {\bf q}) and interband (large {\bf Q}) scattering of the electron pockets of similar strength. This occurs due to similar orbital content of the pockets. Nevertheless, the clear separation of the peaks allows for the purely electronic mechanism of the Cooper-pairing even in the presence of weak or moderate correlations as we show later.  At the same time, much stronger electronic response is found for $n=1.4$ where one clearly finds the strong scattering between electron and hole pockets near ${\bf Q}=(\pi,0)[(0,\pi)]$. It dominates the instability of the electronic gas at this filling arising from the logarithmic divergence of the Lindhard response function near this wave vector due to near nesting of the electron an hole bands separated by {\bf Q}. This is somewhat reminiscent of the electronic structure of the iron-based superconductors and worth studying further.

%
%+++++++++++++++++++++++++++++++   figure   ++++++++++++++++++++++++++++++++++++++++++++++++++
%
%++++++++++++++++++++++++++++++++++++++++++++++++++++++++++++++++++++++++++++++++++++++++

%%%%%%%%%%%%%%%%%%%%%%%%%%%%
\section{Random Phase approximation}
In order to study the superconducting state, % of this compound,
we introduce an intra-site Hubbard model, including intra and inter-orbital interactions ($U$ and $U'$, respectively), as well as the Hund's rule $J$ and pair hopping energy $J'$,
to compute the total  electronic response within the random Phase approximation (RPA) approach. 
In this respect,   the real space picture  of the interacting  Hamiltonian is given by~\cite{Matrins2013} 
%
%--------------------------------------------------------------------
\be
 \label{eq:erl}
 \bl
\mathcal{H}_{int}
=&
 \sum_{\boldsymbol{i}, \nu}
 {U}n_{\boldsymbol{i} \nu \uparrow}n_{\boldsymbol{i} \nu \downarrow} +
\sum_{\boldsymbol{i},\nu\neq\nu'} 
\Big[
({U}'-\frac{J}{2})
 n_{\boldsymbol{i}\nu}n_{\boldsymbol{i} \nu'}  
\\
&
-2 {J} 
\boldsymbol{S}_{\boldsymbol{i} \nu}
\cdot 
\boldsymbol{S}_{\boldsymbol{i} \nu'}
+
{J}'
(
c_{\boldsymbol{i} \nu \uparrow}^\dagger c_{\boldsymbol{i} \nu \downarrow}^\dagger c_{\boldsymbol{i} \nu' \downarrow}c_{\boldsymbol{i} \nu' \uparrow}
+h.c.
)
\Big],
\el
\ee
%------------------------------------------------------------------------------------------------------------------------------------------------------------------------------------------------------------------------------------------------
%
where 
$n_{\boldsymbol{i}\nu}=\sum_{\sigma}  n_{\boldsymbol{i}\nu \sigma}= \sum_{\sigma} c_{\boldsymbol{i} \nu \sigma}^\dagger c_{\boldsymbol{i} \nu \sigma}^{}$,
and 
$\boldsymbol{S}_{\boldsymbol{i} \nu}=\sum_{\sigma,\sigma'} c_{\boldsymbol{i} \nu \sigma}^\dagger   \hat{\sigma}^{}_{\sigma \sigma'} 
 c_{\boldsymbol{i} \nu \sigma'}^{}$.
Therefore, in  the band  representation the total Hamiltonian reads
%
%-------------------------------------------------------------------------
\begin{equation}
\begin{aligned}
\hat{\mathcal{H}}=
&
\sum_{\mu,\boldsymbol{k}}
\epsilon_{\mu}(\boldsymbol{k})
b^{\dagger}_{\mu\boldsymbol{k}}
b^{}_{\mu\boldsymbol{k}}
+
\\
&
\sum_{\substack{\mu\mu' , \boldsymbol{k} \boldsymbol{k'}}}
\Gamma_{\mu\mu'}(\boldsymbol{k},\boldsymbol{k'})
b^{\dagger}_{\mu \boldsymbol{k}}
b^{\dagger}_{\mu,-\boldsymbol{k}} 
b_{\mu'\boldsymbol{k'}}
b_{\mu',-\boldsymbol{k'}},
\end{aligned}
\end{equation}
%-------------------------------------------------------------------------
%
where the quartic terms describe the scattering of pairs $(\boldsymbol{k}\uparrow,-\boldsymbol{k}\downarrow)$ on the pocket $\mu$ to $(\boldsymbol{k'}\uparrow,-\boldsymbol{k'}\downarrow)$ on the pocket $\mu'$, with subsequent scattering amplitude: %. Where	 $\Gamma_{\mu\mu' }$ vertex describes the amplitude of this scattering and it's given by
%------------------------------------------------------------------------------------------------------------------------------------------------------------------------------------------------------------------------------------------------
\be %\nonumber
\begin{aligned}
&
\Gamma_{\mu\mu' }(\boldsymbol{k}_1,\boldsymbol{k}_2)=
\\
&\hspace{0.6cm}
\sum_{ptsq} 
\zeta^{p*}_{-\k_1 \mu} 
\zeta^{t*}_{\k_1 \mu'} 
\left[\Gamma^{sq}_{tp}(\k_1,\k_2)\right]
%\times\\
\zeta^{s}_{\k_2 \mu} 
\zeta^{q}_{-\k_2\mu'} ,
\label{LAHA-scattering}
\end{aligned}
\ee
%------------------------------------------------------------------------------------------------------------------------------------------------------------------------------------------------------------------------------------------------
%
in which, the vertex $\left[\Gamma^{sq}_{tp}(\boldsymbol{k}_1,\boldsymbol{k}_2)\right] $  describes the amplitude of the scattering in orbital basis, and it is a linear combination of the interactions: $U$, $U'$, $J$, and $J'$. 
Here,  $\zeta^{p}_{\k\mu}=\zeta^{\mu *}_{\k p}$ are the matrix elements connecting the orbital and band basis, % in other words,  $\zeta^{p}_{\k\mu}$  
and they represent the orbital contributions to the Fermi surface, see  Fig.~\ref{fig2}(c\&d). \\

The final RPA result of the susceptibility can be addressed  in the form of Dyson equations as follows:
%
%-------------------------------------------------------------------------
\begin{equation}
(\chi_{RPA})^{p,q}_{s,t}=(\chi_{0})^{p,q}_{s,t}+
(\chi_{RPA})^{p,q}_{u,v}(U_{\mu})^{u,v}_{w,z}(\chi_{0})^{w,z}_{s,t},
\end{equation}
%-------------------------------------------------------------------------
%
where $(U_\mu)^{u,v}_{w,z}$ are the coefficients of the interacting Hamiltonian arranged into the three orbital basis matrix, and the repeated indices are summed over. 
For the transversal susceptibility the above equation results in
%
%-------------------------------------------------------------------------
\begin{equation}
(\chi^{\pm}_{RPA})^{p,q}_{s,t}=(\chi^{\pm}_{0})^{p,q}_{s,t}+
(\chi^{\pm}_{RPA})^{p,q}_{u,v}(U^{\pm}_{\mu})^{u,v}_{w,z}(\chi^{\pm}_{0})^{w,z}_{s,t},
\end{equation}
%-------------------------------------------------------------------------
%
where the non-zero components of ${U}_\mu$ are given by
%
%-------------------------------------------------------------------------
\be
\bl\nonumber
&(U^{\pm}_{\mu})^{a,a}_{a,a}=   \tilde{U};\;\;\;\;\;\;
   (U^{\pm}_{\mu})^{a,a}_{b,b}=   \tilde{J};
   \\
&  (U^{\pm}_{\mu})^{a,b}_{a,b}=   \tilde{V};\;\;\;\;\;\;
    (U^{\pm}_{\mu})^{a,b}_{b,a}=   \tilde{J}'.
\el
\ee
%-------------------------------------------------------------------------
however, for the longitudinal susceptibility we find 
\begin{equation}
(\chi^{zz}_{RPA})^{p,q}_{s,t}=\frac{1}{4}
\bigg[
\sum_{\sigma}(\chi^{\sigma\sigma}_{RPA})^{p,q}_{s,t}-\sum_{\sigma\neq{\sigma}'}(\chi^{\sigma{\sigma}'}_{RPA})^{p,q}_{s,t}
\bigg],
\end{equation}
%-------------------------------------------------------------------------
%
where
%
%-------------------------------------------------------------------------
\begin{equation}
(\chi^{\sigma \sigma'}_{RPA})^{p,q}_{s,t}=(\chi^{\sigma \sigma'}_{0})^{p,q}_{s,t}+
(\chi^{\sigma \sigma'}_{RPA})^{p,q}_{u,v}(U^{\sigma'\sigma }_{s})^{u,v}_{w,z}(\chi^{\sigma \sigma'}_{0})^{w,z}_{s,t},
\end{equation}
%-------------------------------------------------------------------------
with
%-------------------------------------------------------------------------
\begin{align*}
&
(U^{\sigma\sigma'}_\mu)^{a,a}_{a,a}=  \tilde{U}
;\;\;\;\ \;\;\;\ 
 (U^{\sigma\sigma'}_\mu)^{a,a}_{b,b}=  \tilde{V};
\\
&
 (U^{\sigma\sigma'}_\mu)^{a,b}_{a,b}=  \tilde{J} 
;\;\;\;\ \;\;\;\ 
  (U^{\sigma\sigma'}_\mu)^{a,b}_{b,a}=  \tilde{J}',
\end{align*}
and finally 
\begin{equation}
(\chi^{\sigma \sigma}_{RPA})^{p,q}_{s,t}=(\chi^{\sigma \sigma}_{0})^{p,q}_{s,t}+
(\chi^{\sigma \sigma}_{RPA})^{p,q}_{u,v}(U^{\sigma \sigma}_{\mu})^{u,v}_{w,z}(\chi^{\sigma \sigma}_{0})^{w,z}_{s,t},
\end{equation}
%-------------------------------------------------------------------------
with
%-------------------------------------------------------------------------
\begin{align*}
(U^{\sigma\sigma}_\mu)^{a,a}_{a,a}= & 0 
;\;\;\;\ \;\;\;\ 
 (U^{\sigma\sigma}_\mu)^{a,a}_{b,b}=  \tilde{V} +\tilde{J};
 \\
(U^{\sigma\sigma}_\mu)^{a,b}_{a,b}= & \tilde{J}+\tilde{V} 
;\;\;\;\ \;\;\;\ 
 (U^{\sigma\sigma}_\mu)^{a,b}_{b,a}=  0.
\end{align*}
%-------------------------------------------------------------------------
%
Despite the fact that the correlations should be moderate in BiS$_2$ systems, we still include the RPA corrections assuming the strength of the interactions to be $U<t$. Nevertheless, as the Lindhard response function shows, there are several intraband and interband scattering peaks, well separated in momentum space. Therefore, we expect them to be enhanced differently once correlations are included in the particle-hole channel. In particular, we see that the RPA spin susceptibilities, despite of moderate interactions, further strengthen the features, present in the Lindhard response function.  For $n=0.2$ the interband and intraband scattering are only slightly enhanced, which indicates very weak instability of the electron gas in the magnetic channel.  The interband and intraband magnetic fluctuations are by far not enough to generate the true magnetic instability. They are, however, sufficiently strong to generate the so-called nodeless $d_{x^2-y^2}$-wave superconductivity with sign change of the gap between electron pockets.  This type of superconductivity was originally proposed for relatively small electron filling within spin fluctuation scenario \cite{Usui2012}. The crucial difference is, however, that this gap possesses strong angular dependence on the Fermi surface sheets making it difficult to identify the global $d_{x^2-y^2}-$wave symmetry.  Furthermore, for large $n=1.4$ the instability with respect to the $(\pi,0)$ or $(0,\pi)$ density wave type ordering is stronger, which is somewhat similar to the iron-based superconductors. Such an instability is known to favor the so-called $s^{\pm}$-wave symmetry of the superconducting order parameter, driven by short-range magnetic fluctuations. We analyze these instabilities quantitatively in the section using the so-called leading angular harmonic approximation, introduced previously\cite{Maiti2011a,Maiti2011b,ahn2014}. As we show later, even in this case there are higher angular harmonics in the superconducting gap on the Fermi surface.

%%%%%%%     Section   %%%%%%%%
\section{Leading angular harmonics approximation}
The LAHA approximation was developed for the iron-based superconductors ~\cite{Maiti2011a,Maiti2011b,ahn2014} as a method to analytically solve and characterize the superconducting gap equation. This method is particularly well suited once the interactions are moderate, which do not give rise to a pronounced enhancement of the spin fluctuations as it is the case in LiFeAs\cite{ahn2014}. In this section we extend it to the BiS$_2$ superconductors. %\\
The basic idea of LAHA is that the $\Gamma_{\mu\mu' }$, defined  in eq.~(\ref{LAHA-scattering}), are  dependent on the angles along different Fermi surface sheets, which are well separated in the BZ. Therefore, they can be defined as simple functions of the momenta $\boldsymbol{k}$ and $\boldsymbol{k'}$ ~\cite{Maiti2011a,Maiti2011b,ahn2014}. In particular, one can decompose the 
$\Gamma_{\mu\mu'}(\boldsymbol{k},\boldsymbol{k'})$
into representations of the tetragonal space group $A_{1g}$ (s-wave) and $B_{1g}$ (d-wave), i.e.,
%------------------------------------------------------------------------------------------------------------------------------------------------------------------------------------------------------------------------------------------------
\begin{equation}\nonumber
\Gamma^{(A_{1g})}(\boldsymbol{k},\boldsymbol{k'})=\Gamma^{s}(\boldsymbol{k},\boldsymbol{k'})=\sum_{m,n}A^{s}_{mn}\Psi^{s}_{m}(\boldsymbol{k})\Psi^{s}_{n}(\boldsymbol{k'}),
\end{equation}
%------------------------------------------------------------------------------------------------------------------------------------------------------------------------------------------------------------------------------------------------
and
%------------------------------------------------------------------------------------------------------------------------------------------------------------------------------------------------------------------------------------------------
\begin{equation}\nonumber
\Gamma^{(B_{1g})}(\boldsymbol{k},\boldsymbol{k'})=\Gamma^{d}(\boldsymbol{k},\boldsymbol{k'})=\sum_{m,n}A^{d}_{mn}\Psi^{d}_{m}(\boldsymbol{k})\Psi^{d}_{n}(\boldsymbol{k'}),
\end{equation}
%-------------------------------------------------------------------------------------------------------- 
with $\Psi^{s}$ and $\Psi^{d}$ being the basis functions of the $A_{1g}$ and $B_{1g}$ representations, respectively. Finally, the $\Gamma^{(A_{1g})}$ and $\Gamma^{(B_{1g})}$ can be expanded to model the angular dependence of the pair-scattering. %\\
In this work, the LAHA approximation will be adapted and implemented to analyze the superconducting instabilities of BiS$_2$ superconductors in two  limiting cases: the small ($n<0.5$), and the large ($n>1.5$) electron fillings.  The Fermi Surface topology is characterized in both cases by the presence of disconnected electron and/or hole pockets centered near the high-symmetry points of the first BZ. Furthermore, we expect the spin singlet state to dominate the actual pseudo-spin-singlet Cooper-pairing and that the Cooper-pairs can be further characterized as even-parity spin singlet solutions. This is due to the fact that the spin-orbit coupling only introduces the overall difference between the longitudinal and the transverse components, which is weakly momentum dependent. Thus, one could consider the Cooper-pairing can be still determined as an even parity wave functions in the psudospin basis. Furthermore, as the admixture of the $p_z$ orbitals is also often weak, the spin remains in most of the cases a good quantum number. 

%%%%%%%%%%%%%%%%%%%%%%%%%%%%%%
\subsection{Low doping limit: $n=0.2$}
The Fermi Surface of the system for small doping levels consists of two electron pockets at $(\pi,0)$, $e_1$, and $(0,\pi)$, $e_2$. In contrast to the Fe-based superconductors, the topology of these pockets is rather non-elliptical (Fig.~\ref{fig2}), implying that higher harmonics (up to $\cos 4 \phi$) need to be considered to correctly describe the scattering processes:
\begin{widetext}
%------------------------------------------------------------------------------------------------------------------------------------------------------------------------------------------------------------------------------------------------
\begin{equation}\label{laha1}
\begin{aligned}
\nonumber
%\
\\
&
\Gamma_{e_{1}e_{2}}^s= U_{ee} \big[1 + 2 \alpha_{ee} (\cos 2 \phi_{k_{e1}} - \cos 2 \phi_{k_{e2}'})
 - 4 \beta_{ee} \cos 2 \phi_{k_{e1}} \cos 2 \phi_{k_{e2}'} 
+ 2 \gamma_{ee} (\cos 4 \phi_{k_{e1}} +  \cos 4 \phi_{k_{e2}'})
\\&\hspace{1cm}
 + 2 \delta_{ee} (\cos 2 \phi_{k_{e1}} \cos 4 \phi_{k_{e2}'} - \cos 2 \phi_{k_{e2}'} \cos 4 \phi_{k_{e1}})
+4 \eta _{ee} \cos 4 \phi_{k_{e1}} \cos 4 \phi_{k_{e2}'} \big] ,
%\
\\
&
\Gamma_{e_{1}e_{2}}^d =
 \tilde{U}_{ee} 
 [-1 - 2 \tilde{\alpha}_{ee} (\cos 2 \phi_{k_{e1}} - \cos 2 \phi_{k_{e2}'})
 + 4 \tilde{\beta}_{ee} \cos 2   \phi_{k_{e1}} \cos 2 \phi_{k_{e2}'} -  2 \tilde{\gamma}_{ee} (\cos 4 \phi_{k_{e1}} +  \cos 4 \phi_{k_{e2}'})
\\&\hspace{1cm}
  \mp 2 \tilde{\delta}_{ee} (\cos 2 \phi_{k_{e1}} \cos 4 \phi_{k_{e2}'} -  \cos 2 \phi_{k_{e1}'} \cos 4 \phi_{k_{e2}})
-4 \tilde{\eta} _{ee} \cos 4 \phi_{k_{e1}} \cos 4 \phi_{k_{e2}'} 
] ,
%\
\\
&
\Gamma^s_{e_i e_i} = U_{ee} \big[1 \pm 2 \alpha_{ee} (\cos 2 \phi_{k_{ei}} + \cos 2 \phi_{k_{ei}'})
 + 4 \beta_{ee} \cos 2 \phi_{k_{ei}} \cos 2 \phi_{k_{e1}'}
+ 2 \gamma_{ee} (\cos 4 \phi_{k_{ei}} + \cos 4 \phi_{k_{ei}'})
\\
&\hspace{1cm}
  \pm 2 \delta_{ee} (\cos 2 \phi_{k_{ei}} \cos 4 \phi_{k_{ei}'} + \cos 2 \phi_{k_{ei}'} \cos 4 \phi_{k_{ei}})
+4 \eta _{ee} \cos 4 \phi_{k_{ei}} \cos 4 \phi_{k_{ei}'} \big] ,
%\
\\
&
\Gamma_{e_{i}e_{i}}^d = \tilde{U}_{ee} 
[ 1 \pm 2 \tilde{\alpha}_{ee} (\cos 2 \phi_{k_{ei}} + \cos 2 \phi_{k_{ei}'})
 + 4 \tilde{\beta}_{ee} \cos 2 \phi_{k_{ei}} \cos 2 \phi_{k_{ei}'} 
+ 2 \tilde{\gamma}_{ee} (\cos 4 \phi_{k_{ei}} + \cos 4 \phi_{k_{ei}'} )
\\&\hspace{1cm}
 \pm 2 \tilde{\delta}_{ee} 
 (
 \cos 2 \phi_{k_{e_i}} \cos 4 \phi_{k_{ei}'} + \cos 2 \phi_{k_{ei}'} \cos 4 \phi_{k_{ei}})
+4 \tilde{\eta} _{ee} \cos 4 \phi_{k_{ei}} \cos 4 \phi_{k_{ei}'} 
].
%\
\end{aligned}
\end{equation}
%------------------------------------------------------------------------------------------------------------------------------------------------------------------------------------------------------------------------------------------------
\end{widetext}
Here the  upper~(lower) sign corresponds to pocket $1~(2)$,  $ \phi_{k_{ei}}$  ($ \phi_{k_{hi}}$ ) represents the angles along the $i^{\rm th}$ electron  (hole) pocket, and
$\Gamma^s$ and $\Gamma^d$ refer to the extended $s$-wave and $d_{x^2-y^2}$-wave Cooper-pairing vertex,  respectively.
As the correlations are weak, it is instructive to consider the LAHA projections not for the RPA but only restricting to the second order perturbation for the interactions. They are shown in Table II. 

%++++++++++++++++++++++++++++++++++++   Table   +++++++++++++++++++++++++++++++++++++++++++++
\begin{table*}
%\scriptsize
\begin{tabular}{|c|c|c|c|c|c|} \hline
\multicolumn{1}{|c|}{$U_{ee}$}  & $\alpha_{ee}$ & $\beta_{ee}$ & $\gamma_{ee}$ & $\delta_{ee}$ & $\eta_{ee}$ \\\hline
\multicolumn{1}{|c|}{0.393} & $-1.463\cdot10^{-4}$ &  $ 1.438\cdot10^{-5}$ &  $8.866\cdot10^{-3}$ & $1.009\cdot10^{-4}$ & $6.172\cdot10^{-4}$  \\\hline
\multicolumn{1}{|c|}{$\tilde{U}_{ee}$}  & $\tilde{\alpha}_{ee}$ & $\tilde{\beta}_{ee}$ & $\tilde{\gamma}_{ee}$ & $\tilde{\delta}_{ee}$ & $\tilde{\eta}_{ee}$ \\\hline
\multicolumn{1}{|c|}{0.041} & $0.0764$ &  $5.912\cdot10^{-3}$ & 0.0516 & $7.885\cdot10^{-3}$ & $2.660\cdot10^{-3}$  \\\hline
\end{tabular}
\caption{Parameters of the LAHA projection of the electronic interactions projected on the extended $s$-wave and $d_{x^2-y^2}$-wave symmetry representations for $n=0.2$ and with $U=0.6$, $V=0.36$ and $J=J'=0.12$ (all in eV).}
\end{table*}
%++++++++++++++++++++++++++++++++++++   Table   +++++++++++++++++++++++++++++++++++++++++++++
%

Observe that the constant part of the interaction in the $A_{1g}$- pairing channel is larger than in the $B_{1g}$-one. Nevertheless, the constant part of the interaction is always repulsive for the Fermi surface topology for small $n$, while the interband part of the repulsion is pair-building for the $B_{1g}$-symmetry, due to the change of the sign of the order parameter between the $(\pi,0)$ and the $(0,\pi)$ pockets. Nevertheless, the inter and intraband repulsions are the same due to the fact that both bands are electron-like. An inclusion of the RPA corrections in the particle-hole channel due to interband nesting helps in achieving $\tilde{U}_{e1e2} > \tilde{U}_{e1e1}, \tilde{U}_{e2e2}$. Nevertheless, all interactions are still very similar to each other as the spin fluctuations are not strongly enhanced, see Fig.3(c). As a result the solution for the superconducting gap acquires a strong angular dependence in the B$_{1g}$-channel in the form
\begin{align}
\begin{split}
\Delta_{e_1}(\phi)&=\Delta_{e} + \bar{\Delta}_e \cos{2\phi}  + \tilde{\Delta}_e \cos{4\phi}	+ \, ...\\
\Delta_{e_2}(\phi)&=-\Delta_{e} + \bar{\Delta}_e \cos{2\phi}  - \tilde{\Delta}_e \cos{4\phi} + \, ...	\end{split}\label{eq:dwaveansatz}
\end{align}
where ... stands for the higher harmonics of $\cos 6 \theta$ and $\cos 8 \theta$, which we have not explicitly included.  Furthermore, it is important to realize that for the intraband and interband repulsive interaction of similar strength, the superconductivity occurs only if the angular dependent harmonics are of the same magnitude as the constant parts of the gap\cite{ahn2014}, which yields nearly nodal behaviour of the superconducting gap on each of the Fermi surface sheets, i.e. we find $\Delta_e \sim \bar{\Delta_e} \sim \tilde{\Delta_e}$. This is in qualitative agreement with the experimental data\cite{Ota2017}. Nevertheless, the most important prediction of the superconducting gap to belong to the $B_{1g}$-irreducible representation is the antiphase character of the constant magnitudes and $\cos{4 \theta}$ harmonics on the two electron pockets, which still needs to be tested experimentally.
Observe, that near-nodal $d_{x^2-y^2}$-wave solution dominates for the Fermi topology consisting of the electron pockets and the angular harmonics are increasing in magnitudes with increasing $n$. Importantly, this is not altered by the inclusion of the spin-orbit coupling and this symmetry appears to be still the most dominant one for $n<0.4$. For larger doping the Fermi surface undergoes a so-called Lifshitz transition and changes its topology\cite{Usui2012,Matrins2013}. Although for $n>0.4$ the unconventional superconducting mechanism cannot be fully excluded, the superconducting mechanism cannot be purely repulsive as the momentum structure of the response function is less pronounced. Therefore it will likely lose against simple $s$-wave symmetry, driven by the electron-phonon interaction.   

%
%+++++++++++++++++++++++++++++++   figure   ++++++++++++++++++++++++++++++++++++++++++++++++++
\begin{figure}[!t]
\label{fig4}
\includegraphics[width=1.0\linewidth]{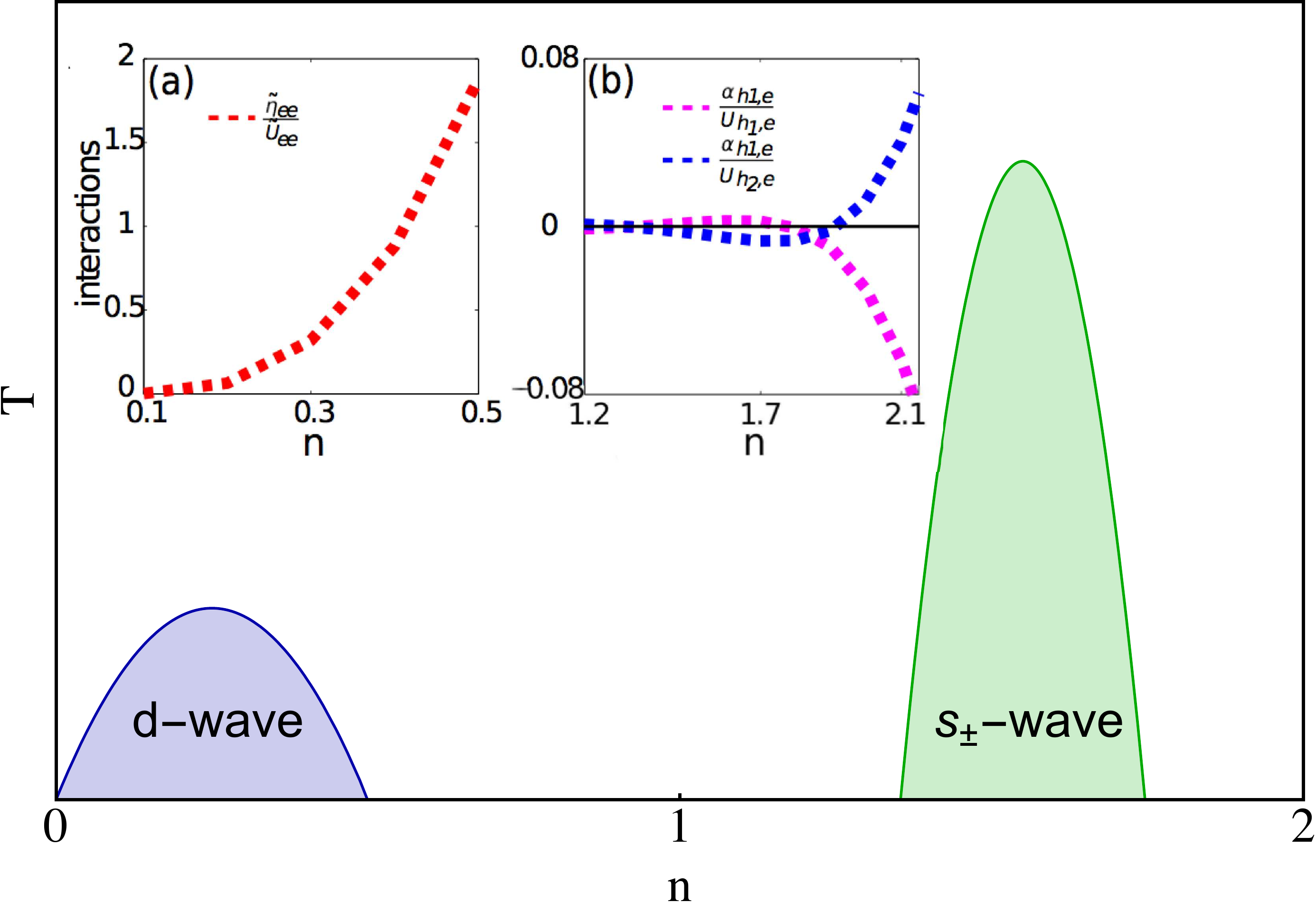}
\caption{(Color online) Putative 
superconducting phase diagram arising from the repulsive interactions. Onset: (a) Evolution of the $\cos 4 \phi_k \cos 4 \phi_{k'}$ angular dependence of the pair-scattering between electron pockets with increasing doping. (b) Evolution of the $\cos 2 \phi_k$ 
angular dependence between electron and hole pockets with increasing doping.
} 
\end{figure}
%++++++++++++++++++++++++++++++++++++++++++++++++++++++++++++++++++++++++++++++++++++++++++

%%%%%%%%%%%%%%%%%%%%%%%%%%%
\subsection{Intermediate doping limit: $n=1.4$}
For a larger doping level $n$, we find a richer and more complicated Fermi Surface: the electron pockets described in the previous section have now merged giving rise to two hole pockets located at $\Gamma$ and $M$. On the other hand, elliptical electron pockets from the upper energy band appear at $X$ and $Y$. All pockets have a similar radius, and are separated by a large  wave vector ${\bf Q}=(\pi,0)$ or $(0,\pi)$. 

%++++++++++++++++++++++++++++++++++++   Table   +++++++++++++++++++++++++++++++++++++++++++++
\begin{table*}
	\begin{tabular}{|c|c|c||c|c|c||c|c|c|c|c|c|c|} \hline
		\multicolumn{1}{|c|}{$U_{ee}$} & $\alpha_{ee}$ & $\beta_{ee}$  & $U_{h1h1}$ & $U_{h2h2}$ & $U_{h1h2}$ & $U_{h1e}$ & $U_{h2e}$ & $\alpha_{h1e}$ & $\alpha_{h2e}$ \\\hline
		\multicolumn{1}{|c|}{0.405} & $-7.909\cdot10^{-4}$ &  $3.156\cdot10^{-5}$ & 0.303 & 0.321  & 0.309 & 0.237 & 0.238
		& $9.928\cdot10^{-4}$ & $-2.806\cdot10^{-3}$  \\\hline
		\multicolumn{1}{|c|}{$\tilde{U}_{ee}$}  & $\tilde{\alpha}_{ee}$ & $\tilde{\beta}_{ee}$  & $\tilde{U}_{h1h1}$ & $\tilde{U}_{h2h2}$ & $\tilde{U}_{h1h2}$ & $\tilde{U}_{h1e}$ & $\tilde{U}_{h2e}$ & $\tilde{\alpha}_{h1e}$ & $\tilde{\alpha}_{h2e}$  \\\hline
		\multicolumn{1}{|c|}{0.0262} & 0.0894 &  $8.144\cdot10^{-3}$ & 0.0187 & $2.644\cdot10^{-3}$ & $6.816\cdot10^{-3}$  & 0.0221 &  $8.143\cdot10^{-3}$  & 0.179 & 0.182 \\\hline
	\end{tabular}
	\caption{Parameters of the LAHA projection of the electronic interactions projected on the extended $s$-wave and $d_{x^2-y^2}$-wave symmetry representations for $n=1.4$ with $U=0.6$, $V=0.36$ and $J=J'=0.12$ (all in eV). }
\end{table*}
%++++++++++++++++++++++++++++++++++++   Table   +++++++++++++++++++++++++++++++++++++++++++++

Due to the presence of hole pockets, we need to take into account hole-hole $\Gamma_{h_i h_j}$, and electron-hole $\Gamma_{e_i h_j}$ scattering vertices. On the other hand, all four pockets are highly symmetric now. Therefore, we neglect the harmonics higher than $\cos 2 \phi$ on the electron pockets for simplicity:
%------------------------------------------------------------------------------------------------------------------------------------------------------------------------------------------------------------------------------------------------
\begin{equation}
\nonumber
\begin{aligned}
\Gamma^s_{e_i e_i} = U_{ee} [
1 \pm 2 \alpha_{ee} (\cos 2 \phi_{k_{ei}} + \cos 2 \phi_{k_{ei}'})\\
 + 4 \beta_{ee} \cos 2 \phi_{k_{ei}} \cos 2 \phi_{k_{ei}'}
 ],
\end{aligned}
\end{equation}
%------------------------------------------------------------------------------------------------------------------------------------------------------------------------------------------------------------------------------------------------
and 
%------------------------------------------------------------------------------------------------------------------------------------------------------------------------------------------------------------------------------------------------
\begin{equation}
\nonumber
\begin{aligned}
\Gamma^d_{e_i e_i}=\tilde{U}_{ee} 
[1 \pm 2 \tilde{\alpha}_{ee} (\cos 2 \phi_{k_{ei}} + \cos 2 \phi_{k_{ei}'})\\
 + 4 \tilde{\beta}_{ee} \cos 2 \phi_{k_{ei}} \cos 2 \phi_{k_{ei}'}
 ].
\end{aligned}
\end{equation}
%------------------------------------------------------------------------------------------------------------------------------------------------------------------------------------------------------------------------------------------------
\normalsize
and the hole pockets can be reasonably well approximated by circle i.e. by a constant term. Therefore we have,
 %------------------------------------------------------------------------------------------
\begin{equation}
\begin{aligned}
\nonumber
\Gamma_{h_i h_i}
=
&
\Gamma^s_{h_i h_i}+\Gamma^d_{h_i h_i} =
 U_{h_i h_i}+\tilde{U}_{h_i h_i} \cos 2 \phi_{k_{hi}} \cos 2 \phi_{k_{hj}'}
 ,
\\
\nonumber
\Gamma_{h_{1}h_{2}}
=
&
\Gamma_{h_{1}h_{2}}^s +\Gamma_{h_{1}h_{2}}^d
\\
=&
-U_{h_1 h_2}-\tilde{U}_{h_1 h_2} \cos 2   \phi_{k_{hi}} \cos 2 \phi_{k_{hj}'},
\end{aligned}
\end{equation}
%------------------------------------------------------------------------------------------
Finally, the electron-hole vertices read
%------------------------------------------------------------------------------------------
\begin{equation}\label{laha1.1}
\begin{aligned}
\nonumber
\Gamma_{h_i e_j}=\Gamma^s_{h_i e_j}+\Gamma^d_{h_i e_j} = \pm U_{h_i e_j}
\Big[
1 \pm \alpha_{h_i e_j}\cos(2\phi_{k_{ej}}')
\Big]
\\
+\tilde{U}_{h_i e_j}\cos(2\phi_{k_{hi}})
\Big[ 
1 \pm \tilde{\alpha}_{h_i e_j}\cos(2\phi_{k_{ej}}')
\Big],
\end{aligned}
\end{equation}
%------------------------------------------------------------------------------------------
here, upper and lower sign corresponds to $\Gamma_{h_1 e_1}$ and $\Gamma_{h_2 e_2}$ respectively.
We present the results of the projections for the bare interactions in Table III.  As  one clearly sees, the largest repulsive interaction in the Cooper-pairing channel occurs in the $A_{1g}-$channel. However, in contrast to the case of smaller $n$, here the Fermi surface topology allows to have the $A_{1g}$-solution due to the appearance of the hole pockets near the $\Gamma$ and M points of the Brillouin Zone. As a result, the repulsion between the electron and the hole pockets becomes effectively pair-building for the so-called $s^{\pm}$-wave symmetry due to the change of sign of the superconducting gap between electron and hole pockets. Furthermore, due to the different character of the bands, the interaction between electron and hole pockets will be enhanced even for the moderate renormalization within RPA, as seen from Fig.3(d). Therefore, this state wins over $d_{x^2-y^2}$-wave for these doping concentrations. In particular, we find for the superconducting on the electron and hole pockets in the $A_{1g}$-channel
\begin{align}
\begin{split}
\Delta_{h_1}(\phi)&=\Delta_{h_1}	\\
\Delta_{h_2}(\phi)&=\Delta_{h_2}	\\
\Delta_{h_3}(\phi)&=\Delta_{h_3}	\\
\Delta_{e_1}(\theta)&=\Delta_{e}+\bar{\Delta}_{e}\cos{2\theta}	\\
\Delta_{e_2}(\theta)&=\Delta_{e}-\bar{\Delta}_{e}\cos{2\theta}	\\
\end{split}\label{eq:swaveansatz}
\end{align}
where the signs of the constant gap on the electron and hole pockets are opposite. It is interesting to note that even for the moderate correlations within RPA the interband repulsion between the electron and the hole pockets becomes larger than the intraband or the interband within pockets of the same character, i.e. $h_1 h_2$ or $e_1 e_2$. As a result the gaps on the hole pockets can be considered as constants to a good approximation. On the electron pockets the gaps still acquire some significant angle dependence with $\bar{\Delta}_e < \Delta_e$. 
Analysing further the behaviour of the interactions and solving the linear version of the BCS gap equation we found that this part of the doping phase diagram is dominated by the $s^{+-}-$wave symmetry with sign changing gap on the electron and hole pockets. Although such a filling factor was not achieved for $BiS_2$-systems, yet it would be interesting to investigate its potential realization in these and similar systems.

%%%%%%%%%%%%%%%%%%%%%%%%%%%%%%%%%%%%%%%%%%%%%%%%%%%%%%%%%%%%%%%%%
\section{Conclusions}  

To conclude, we develop the effective low-energy Hamiltonian of BiS$_2$ lyered superconductors based on the fully relativistic ab-initio band structure calculations, including the spin-orbit coupling, relevant for Bi-based systems. The model consists of the tight-binding Hamiltonian, based on all three $p$-orbitals, $p_x$, $p_y$, and $p_z$. The spin orbit coupling introduces weak spin anisotropy, which does not modify the spin structure of the Cooper-pairing. In particular, due to relatively weak contribution of the $p_z$ orbitals to the Fermi surface, the Cooper-pairing can be still regarded as even parity spin singlet state.  Despite of the weakness of correlations for the $p-$electrons we find that the purely repulsive interaction still yields unconventional superconductivity as soon as the interand and intraband interactions can be separated in the momentum space. In particular, for $n\sim 0.3$ the superconducting gap possesses the global $d_{x^2-y^2}-$wave symmetry yet it acquires strong angular dependence forming accidental nodes or deep minima on the electron pockets, located at X and Y points of the BZ. This generally agrees with recent ARPES experiments\cite{Ota2017} and indicates that the mechanism of the Cooper-pairing is still repulsive in nature. Despite the near-nodal or nodal behavior and strong angular dependence of the gaps on the Fermi surface, the global symmetry remains $d_{x^2-y^2}$-wave. This is reflected by the opposite signs of the $\cos 4\phi$ harmonics on the two electron pockets, respectively.  This interesting predication needs to be further verified experimentally. Furthermore, with increasing $n$ we find another doping region where the interaction has a well-defined momentum structure with a clear separation of the inetrband and intraband scattering. The topology of the Fermi surface is then similar to the case of iron-based superconductors with electron and hole pockets. In this case the nesting between electron and hole bands promotes the nodeless $A_{1g}$-symmetry representation to be the dominant solution in this case even for the weakly correlated case.

%%%%%%%%%%%%%%%%%%%%%%%%%%%%%%%%%%%%%%%%%%%%%%%%%%%%%%%%%%%%%%%%%
\section*{Acknowledgements}  
We acknowledge helpful discussions with F. Ahn, D. Altenfeld, A. Donkov and J. Knolle. 
S.C. and  A.A. wish to acknowledge the Korea Ministry of Education, Science and Technology, Gyeongsangbuk-Do and Pohang City for Independent Junior Research Groups at the Asia Pacific Center for Theoretical Physics. 
The work by  A.A. was supported through  NRF funded by MSIP of Korea (2015R1C1A1A01052411), and by  Max Planck POSTECH / KOREA Research Initiative (No. 2011-0031558) programs through NRF funded by MSIP of Korea. I.E. acknowledges support by the Ministry of Education
and Science of the Russian Federation in the framework of Increase Competitiveness
Program of NUST MISiS (K2-2016-067).
%%%%%%%%%%%%%%%%%%%%%%%%%%%%%%%%%%%%%%%%%%%%%%%%%%%%%%%%%%%%%%%%%

%%%%%%%%%%%%%%%%%%%%%%%%%%%%%      References        %%%%%%%%%%%%%%%%%
%%%%%%%%%%%%%%%%%%%%%%%%%%    \bibliography  %%%%%%%%%%%%%%%%%%%%%%%%
%\section*{References}
%\bibliographystyle{PRB}

\bibliography{References}
%%%%%%%%%%%%%%%%%%%%%%%%%%%%%      References        %%%%%%%%%%%%%%%%%
%%%%%%%%%%%%%%%%%%%%%%%%%%%%%      References        %%%%%%%%%%%%%%%%%

\end{document}